\documentclass{elsart}
\usepackage{natbib}
\usepackage{graphicx}
\usepackage{epsfig}

\begin{document}
\runauthor{Deligny et al.}
\begin{frontmatter}
\title{Angular Power Spectrum Estimation of Cosmic Ray Anisotropies with
Full or Partial Sky Coverage}
\author[PCC,APC]{O. Deligny}
\author[PCC,GRECO,APC]{E. Armengaud}
\author[PCC,APC]{T. Beau}
\author[LPNHE]{P. Da~Silva}
\author[LPNHE]{J.-Ch. Hamilton}
\author[PCC,APC]{C. Lachaud}
\author[LPNHE]{A. Letessier-Selvon}
\author[GRECO,APC]{B. Revenu}

\address[LPNHE]{LPNHE (Universit\'es Paris 6 \& 7, CNRS-IN2P3), Paris,
France}
\address[PCC]{PCC-Coll\`ege de France (CNRS-IN2P3), Paris, France}
\address[GRECO]{IAP-GRECO, Paris, France}
\address[APC]{APC, (Universit\'e Paris  7, CNRS-IN2P3), Paris, France}

\begin{abstract}
We study the angular power spectrum estimate in order to search for
large scale ani\-so\-tro\-pies in the arrival directions distribution
of the highest-energy cosmic rays. We show that this estimate can
be performed even in the case of partial sky coverage and validated over the full sky
under the assumption that the observed fluctuations are statistically spatial stationary. 
If this hypothesis - which can be tested directly on the data - is not satisfied,
it would prove, of course, that the cosmic ray sky is non isotropic but also that the
power spectrum is not an appropriate tool to represent its anisotropies, whatever 
the sky coverage available. We apply the method to simulations
of the Pierre Auger Observatory, reconstructing an input power spectrum
with the Southern site only and with both Northern and Southern ones.
Finally, we show the improvement that a full-sky observatory 
brings to test an isotropic distribution, and we discuss the 
sensitivity of the Pierre Auger Observatory to large scale anisotropies.
\end{abstract}
\end{frontmatter}

\section{Introduction}

The origin of the highest energy cosmic rays is a theoretical
challenge of modern astrophysics, and is subject of much experimental
efforts.  Above $10^{20}$ eV, the current data are too scarce for one
to make any definitive statement about the existence or the lack of
the GZK cutoff, as well as a statistically meaningful information
about the arrival direction distribution. Whereas the
AGASA experiment ~\cite{agasa} is over since January 2004, a new
generation of experiments especially dedicated to the highest energies
is emerging, and the first of them is the Pierre Auger Observatory
currently under construction~\cite{auger}. For a recent review on the
state of the art of the highest-energy cosmic rays, we refer the
reader to \cite{cronin} for instance.

The distribution of the arrival directions is certainly one of the
most crucial observable in order to yield some evidences about the
sources of the highest-energy cosmic
rays~\cite{isola,sigl2,sigl}. When trying to point out large scale
anisotropies, one is naturally led to work with the angular power
spectrum of the arrival direction distribution. The detection of large
scale anisotropy could probe certain classes of sources and/or test
certain propagation models in presence of magnetic fields to be
associated with such large scale celestial patterns. In addition, the
evidence for large scale anisotropy around 1~EeV\footnote{1 
EeV$\equiv 10^{18}$ eV} claimed by the AGASA
collaboration~\cite{anisagasa2,anisagasa} motivates even more this kind of
studies.

From the PeV energy range, the flux of cosmic rays is so low that we
need ground based experiments with large collecting areas measuring
the secondary products of the interaction of the cosmic ray in the
upper atmosphere. As any ground based experiment has only at one's
disposal a limited field of view in declination distribution, anisotropy 
analysis are generally done owing to the nearly uniform exposure in 
right ascension by using a 1-dimensional coordinate system instead of 
the natural 2-dimensional one over the sphere. This is the case for 
example of the Rayleigh formalism~\cite{linsley} which necessarily 
corrupts the sensitivity to tiny anisotropies. 
In order to exploit the angular power spectrum analysis methods, it is
assumed within the cosmic rays community that a full
exposure of the sky is required \cite{sommers,anchordoqui}. The aim of
this paper is to show that this conclusion arises only because of the
choice of the spherical harmonic coefficients estimate, and to show
that with another choice of estimate, standard anisotropy analysis
methods can be used even with a partial and non-uniform coverage of the 
celestial sphere. 

By denoting $\vec{n}_i$ each cosmic ray arrival 
direction, the standard estimate of the spherical harmonic coefficients 
is computed through \[ a_{\ell m} = \frac{1}{\mathcal{A}}\sum_{i=1}^N 
\frac{Y_{\ell m}(\vec{n}_i)}{\omega(\vec{n}_i)}
\]
where $N$ is the total number of events, $\omega(\vec{n})$ is the 
relative exposure function of the considered experiment, and $\mathcal{A}$ 
a normalization constant taken as $\sum_{i=1}^N 1/\omega(\vec{n}_i)$. As well
known, the use of $1/\omega$ allows for decoupling the modes when working 
with a variable exposure over the whole celestial sphere, but breaks 
down in case of partial exposure of the sky, because it is no longer 
possible to perform the full sky integrations that are required to measure 
the multi-poles of the celestial cosmic ray intensity \cite{sommers}. 

In this paper, we choose to introduce and adapt the quadratic estimator
method that is widely used in the Cosmic Microwave Background  analysis 
(see e.g.~\cite{hivon}) where effects of a partial exposure can be deconvoluted 
from the observations in order to recover the true underlying power spectrum.
We show that the application of this method allows for the
standard anisotropy analysis with an exposure possibly going to zero
in some parts of the sky. This point is of major interest
for most cosmic rays experiments, as the Southern site of the Auger
Observatory for instance.

Another approach to power spectrum estimation is through maximum
likelihood (see ~\cite{bjk,borrill} and~\cite{cras_jch} for a review)
which has the advantage of solving exactly the problem.
It however requires an explicit representation of the sky covariance and
is computationally very time consuming and numerically hard to achieve
on large datasets. The quadratic estimate proposed here avoids this difficulty 
using a Monte-Carlo simulation (this is why it is often refered to as a
``frequentist approach''). 

This paper is organized as follows~: in the next section, we describe our method and
compute all the statistical properties of our choice of angular power spectrum
estimates. From section 3 and on we apply it to the forthcoming Auger observatory. 
In section 3, we present the relevant informations about this experiment that we
need in the context of angular power spectrum estimation. In
section 4, we discuss the constraints that the Auger
Observatory can put on isotropic distribution of cosmic rays
at ultra-high energy. At last, in section 5, we extend the
analysis to the case of a large scale anisotropic distribution.

\section{Angular power spectrum with a partial sky coverage}

\subsection{Generalities}

The number of cosmic rays observed per unit solid angle 
$\mathrm{d}N/\mathrm{d}\Omega$ is a Poisson random
variable in each direction $\vec{n}$, whereas considered
as a function of $\vec{n}$, this is a Poisson random process.
We model it with the two dimensional quantity~: 
\[
\frac{\mathrm{d}N}{\mathrm{d}\Omega}\left(\vec{n}\right)=
\mathcal{N}(\vec{n})=\sum_{i=1}^{N} \delta(\vec{n},\vec{n}_i)
\]
where $\delta$ is the Dirac delta function on the surface of the unit
sphere, and $\vec{n}_i$ the position of the $i^{th}$ cosmic ray.
The total number of cosmic rays observed is then 
$\int\mathcal{N}(\vec{n})\mathrm{d}\vec{n}=N$. This distribution follows
a Poisson law $\mathcal{P}(\nu(\vec{n}))$ with an averaged intensity density
in the direction $\vec{n}$~: 
\[
\nu(\vec{n}) = \frac{N}{4\pi f_1}W(\vec{n})\,(1+\Delta(\vec{n}))
\]
where $W$ is the relative coverage of the experiment 
varying from 0 to 1, $f_1=\frac{1}{4\pi}\int W(\vec{n})\mathrm{d}\vec{n}$
the fraction of the sky effectively covered by the experiment, and
$\Delta$ some continuous stochastic field that measures the departure 
from isotropy. The stochastic field $\Delta$ is assumed to have a 
zero expectation value~: 
\[
\left<\Delta(\vec{n})\right>_r = 0
\]
where we have introduced the average over all the possible realizations 
of the random phases of the $\Delta$ field. The expansion of $\Delta$ on 
the spherical harmonics basis is given by~:
\[
\Delta(\vec{n}) = \sum_{\ell\ge 0}\sum_{m=-\ell}^{m=\ell} a_{\ell m} Y_{\ell m}(\vec{n})
\]

It is natural to introduce the $\Delta$ random field  because of the probable 
stochastic nature of the cosmic ray sources distribution and
propagation through magnetic fields.  The possible
anisotropies we want to characterize are \emph{a priori} of random nature. 
In particular, many theoretical models of UHECRs anisotropies are based on at least 
partly random configurations of sources and magnetic fields.
We are driven to interpret a particular set of events as one specific 
realization of the random process; and we are led to characterize 
the underlying random process properties through the angular power spectrum of the
data of the only set of events we have.

Turning now to the two point correlation function of the $\Delta$ field,
we assume this function to be only dependent on the angular distance
between two points on the sphere~:
\[
\left<\Delta(\vec{n})\Delta^{\star}(\vec{n}^{\prime})\right>_r
\equiv \xi(\vec{n}\cdot\vec{n}^{\prime})
\]
This is a very strong and important hypothesis as it is the basis
of any power spectrum estimation, in the classical sense we want
to give to it. In particular,
this assumes that the statistical properties of $\Delta$ are the same 
over the whole celestial sphere. In the following, we will refer to
this hypothesis as \emph{spatial stationarity} by analogy 
with a stationary time dependent problem, where the notion of stationarity 
is used when the two point correlation function depends only on
the time difference\footnote{We use here the vocabulary of ``spatial
stationarity'' rather than the one of ``homogeneity'' by arbitrary choice.}.
Making a simple expansion of $\xi(\vec{n}\cdot\vec{n}^{\prime})$ 
onto the Legendre polynomials\footnote{We use here the spherical
harmonics addition theorem~:\[ \sum_{m=-\ell}^{\ell} Y_{\ell m}
(\vec{n})Y^{\star}_{\ell m}(\vec{n}^{\prime}) = \frac{2\ell+1}{4\pi}
P_\ell(\vec{n}\cdot\vec{n}^{\prime})\]}~:
\[
\xi(\vec{n}\cdot\vec{n}^{\prime}) = \sum_{\ell} \frac{2\ell+1}{4\pi}
C_\ell P_\ell(\vec{n}\cdot\vec{n}^{\prime}) = \sum_{\ell,m}
C_\ell Y_{\ell m}(\vec{n})Y^{\star}_{\ell m}(\vec{n}^{\prime})
\]
and of $\left<\Delta(\vec{n})\Delta^{\star}(\vec{n}^{\prime})\right>_r$ 
onto the spherical harmonics~:
\[
\left<\Delta(\vec{n})\Delta^{\star}(\vec{n}^{\prime})\right>_r =
 \sum_{{\ell_1},m_1}
 \sum_{{\ell_2},m_2}
\left<a_{{\ell_1} m_1}a^{\star}_{{\ell_2}m_2}\right>_r
Y_{{\ell_1} m_1}(\vec{n})Y^{\star}_{{\ell_2} m_2}(\vec{n}^{\prime})
\]
we clearly see by identification that spatial stationarity leads to a diagonal covariance matrix of the $a_{{\ell} m}$
coefficients~:
\[
\left<a_{{\ell_1} m_1}a^{\star}_{{\ell_2}m_2}\right>_r =
C_{\ell_1}\delta_{{\ell_1}{\ell_2}}\delta_{m_1m_2}
\]
where $C_{\ell}$ is the angular power spectrum of the fluctuations.

The angular power spectrum is therefore a two point correlation function 
in $\ell$ space. It gives information on the correlation between two
angular directions separated by an angular scale $\simeq 1/\ell$ (in
radians). For a Gaussian field $\Delta$, the $C_\ell$ power spectrum
characterizes completely the fluctuations of the field as the even
order moments are obtained from the second order moment and the odd
order moments are zero (Wick's theorem). In the general case, this is
no longer true and higher order moments
(three points correlation function and so on ...) are necessary in order to fully
characterize the field. One should note however, that, if the observed
comic ray sky is the result of a sequence of many random processes and 
assuming no single process dominates (e.g. the sky is not the result of a single 
dominant source with no magnetic field), the fluctuations will be of Gaussian nature 
according to the central limit theorem. In any case, higher order moments can be computed
on the data and departure from a Gaussian behavior can be measured. 

As outlined before, the data set we are dealing with is a 
{\em Poisson sample} of the random field $\Delta$. Consequently, we have to introduce 
a second kind of average~: the average over all possible 
sample configurations $\left<\cdot\right>_P$. 
Therefore, from now on, we use the notation  $\left<\cdot\right>
\equiv \left<\,\left<\cdot\right>_P\right>_r$ to express this double
average over the possible configurations of $N$ events 
and over the possible realizations of the $\Delta$ random field. 
From the elementary Poisson statistic properties, it is easy to 
show that~:
\[
\left<\mathcal{N}(\vec{n})\right> = \left<\left<\mathcal{N}(\vec{n})\right>_P\right>_r\, 
= \,\left<\nu(\vec{n})\right>_r\, = \,\frac{N}{4\pi f_1}W(\vec{n})
\]
and~:
\begin{eqnarray*}
\left<\mathcal{N}(\vec{n})\mathcal{N}(\vec{n}^{\prime})\right> &=&
\left(\frac{N}{4\pi f_1}\right)^2 W(\vec{n})W(\vec{n^{\prime}})
(1+\left<\Delta(\vec{n})\Delta(\vec{n}^{\prime})\right>_r) \\
&&+\, \frac{N}{4\pi f_1}W(\vec{n}) \delta(\vec{n},\vec{n}^{\prime})
\end{eqnarray*}

\subsection{Definition of the spherical harmonic coefficients estimate}

We want to build an estimate of the harmonic expansion coefficients 
of $\Delta$. In the cases we are interested in, the field $\Delta$ 
is not measured uniformly over the whole celestial sphere. This is 
due to the non uniform exposure of cosmic ray experiments. 
For a single experiment, the knowledge is even limited to a given 
region in the sky and no information on $\Delta$ is available 
elsewhere. Moreover, in this given region, the exposure is not 
uniform and generally depends on declination. When combining data 
from two observatories, the exposure becomes full sky but non
uniform. This is shown for instance with Sugar and AGASA coverage
in~\cite{anchordoqui}, or with Auger Southern and Northern sites
in~\cite{sommers}. 

All these configurations can be described through the 
introduction of the window field $W(\vec{n})$ that measures
the relative exposure in the direction $\vec{n}$ on the sky. 
This field can even vanish in some regions. 
Thus, $\tilde{\Delta}(\vec{n})=\Delta(\vec{n})\times W(\vec{n})$
is the quantity we have access to experimentally and not simply
$\Delta(\vec{n})$ as in the case of a uniform 
and full sky coverage. This has an immediate effect in the 
$C_\ell$ determination as we cannot compute the expansion 
of the field we intended to. We only have access to 
what is called the pseudo-power spectrum $\tilde{C}_\ell$ of the
product of the two fields. A simple way to go back to the true
$C_\ell$  from the measurement of $\tilde{C}_\ell$ was proposed for
Cosmic Microwave Background analysis by~\cite{hivon} and has been
widely used in this community for various 
experiments~\cite{boomerang,archeops,wmap}. We will soon show that
the convolution kernel which mixes the modes of the angular
spectrum we want to measure is the same as the one found in
the framework of the CMB. 

We denote our estimates $\tilde{a}_{\ell m}$ and
we define them as~:
\[
\tilde{a}_{\ell m} = \int_{4\pi} \mathrm{d}\vec{n} \,Y_{\ell m}^{\star}(\vec{n}) 
\frac{\mathcal{N}(\vec{n})-\frac{N}{4\pi f_1}W(\vec{n})}{\frac{N}{4\pi f_1}}
\]

Clearly, $\left<\tilde{a}_{\ell m}\right>=0$, as well as for
the expectation value of the true coefficients.

\subsection{The bias on the angular power spectrum estimate}

For reasons that will soon become clear, we introduce the
following coupling kernel as in \cite{hivon}~:
\[
K_{\ell m \ell^{\prime} m^{\prime}}=\sum_{\ell_1 m_1} w_{\ell_1 m_1}
\int_{4\pi} \mathrm{d} \vec{n} \,Y_{\ell^{\prime} m^{\prime}}(\vec{n}) 
Y^\star_{\ell m}(\vec{n}) Y_{\ell_1 m_1}(\vec{n})
\]
where we have expanded the window field on the spherical harmonics
basis. Let us also introduce the power spectrum of the window field~:
\[
\mathcal{W}_{\ell}=\frac{1}{2\ell+1}\sum_{m=-\ell}^{\ell} \left|w_{\ell m}\right|^2
\]
Turning now to the correlation between two multi-pole estimates, it
is easy to show that~:
\[
<\tilde{a}_{\ell m} \tilde{a}^\star_{\ell^\prime m^\prime}> =
\sum_{\ell_1 m_1} C_{\ell_1} K_{ \ell m \ell_1 m_1}K^\star_{\ell^\prime
m^\prime \ell_1 m_1 }+\frac{4\pi f_1}{N}K_{\ell m \ell^\prime m^\prime}
\]

We then estimate the power spectrum
$\tilde{C}_{\ell}$ simply by taking the empiric average over $m$~:
\[	
\tilde{C}_\ell = \frac{1}{2\ell+1}\sum_{m=-\ell}^{m=\ell} |\tilde{a}_{\ell m} | ^2
\]
This yields to~:
\[
\left<\tilde{C_\ell}\right> = \frac{1}{2\ell+1}\sum_{m=-\ell}^{\ell}\bigg(\sum_{\ell_1 m_1}C_{\ell_1}
\left|K_{\ell m \ell_1 m_1}\right|^2+\frac{4\pi f_1}{N}K_{\ell m \ell m} \bigg)
\]
In \cite{hivon}, it has been shown that the first term is equivalent
to a mode-mode coupling matrix $M_{\ell\ell_1}$~:
\[
\left<\tilde{C_\ell}\right> = \sum_{\ell_1}M_{\ell\ell_1}C_{\ell_1}
+\frac{4\pi f_1}{N}\frac{1}{2\ell+1}\sum_{m=-\ell}^{\ell}K_{\ell m \ell m}
\]
where the $M_{\ell\ell_1}$ matrix elements are~:
\[
M_{\ell \ell_1}=\frac{2\ell_1+1}{4\pi}\sum_{\ell_2}\left(2\ell_2+1\right)
\mathcal{W}_{\ell_2}\left(\begin{array}{ccc}\ell &\ell_1 &\ell_2 \\ 0 &0
&0\end{array}\right)^2 
\]
which makes use of the Wigner 3-j symbols. By expanding the second
term onto the Wigner 3-j symbols, and after some manipulations, it
is easy to show that~:
\[
\frac{1}{2\ell+1}\sum_{m=-\ell}^{\ell}K_{\ell m \ell m} =
\frac{w_{00}}{\sqrt{4\pi}} = f_1
\]
leading to~:
\[
\left<\tilde{C_\ell}\right> = \sum_{\ell_1}M_{\ell\ell_1}C_{\ell_1}+\frac{4\pi f_1^2}{N}
\]
We therefore have a simple and analytical link between our estimate
and the true $C_\ell$ for a sky observed with a varying and/or incomplete
exposure. Apart from a bias, our estimate is just the convolution of
the true power spectrum by a kernel whose properties can be determined
analytically from the shape of the window.

At last, in~\cite{hivon}, it is shown that the effect of $M$ on
a constant is a multiplication by the second moment of the window
$f_2=\frac{1}{4\pi}\int W^2(\vec{n})\mathrm{d}\vec{n}=\sum_\ell 
\frac{2\ell+1}{4\pi}\mathcal{W}_\ell$. Therefore, we can go back
to the angular power spectrum of the $\Delta$ field through~:
\[
\left<C^{\mathrm{exp}}_\ell\right>=\sum_{\ell^\prime}M^{-1}_{\ell \ell^\prime}
\left<\tilde{C}_{\ell^\prime}\right>
= C_{\ell}+\frac{4\pi}{N}\frac{f_1^2}{f_2}
\]
We see that the experimental power spectrum is unmixed and
asymptotically unbiased. The bias term can be easily computed
analytically and is purely induced by the finite number of arrival
directions that are available, that is, purely induced by the Poisson
statistics of $\mathcal{N}$.

\subsection{The variance of the angular power spectrum estimate}

From the fourth moment of $\mathcal{N}$ and the Wick's theorem,
there is no difficulty to compute the correlation between four multi-poles
estimates. However, this calculation is rather long and tedious, so
we don't reproduce it in details. As in cosmic ray physics, the null
hypothesis we want to test is isotropy, we are interested in 
$C_{\ell}=0$. In this case, the result for the covariance on 
$\tilde{C}_\ell$ is found to be~:
\[
\mathrm{Cov}(\tilde{C}_\ell,\tilde{C}_{\ell^\prime}) =
\bigg(\frac{4\pi f_1}{N}\bigg)^2 \frac{2\pi}{2\ell^\prime+1}M_{\ell\ell^\prime}
\]
Therefore, the variance on the experimental power spectrum 
simply reads~:
\[
V(C_{\ell}^{\mathrm{exp}}) = \sum_{\ell_1,\ell_2}M^{-1}_{\ell \ell_1}
\mathrm{Cov}(\tilde{C}_{\ell_1},\tilde{C}_{\ell_2})(M^{-1}_{\ell_2 \ell})^T
=\bigg(\frac{4\pi f_1}{N}\bigg)^2 \frac{2\pi}{2\ell+1}M_{\ell\ell}^{-1}
\]

\begin{figure}[!ht]
\centering\epsfig{file=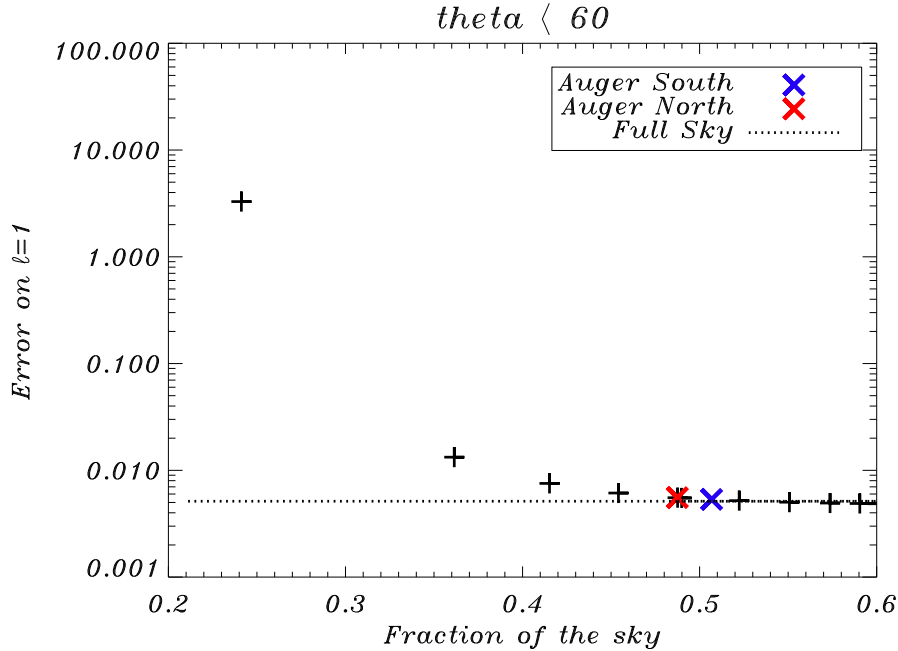,width=12cm,height=9cm}
\caption{Error bars of the reconstructed dipole as a function of covered
portion of the sky. For an experiment covering more than 40\% of the
sky, the error bars become stable and the corresponding dipole estimate
makes sense. For values smaller than 40\%, the $M_{\ell_{1}\ell_{2}}$ 
matrix is no longer regular and, as a result, the error bars explode. This
conclusion arises with the conservative choice of 
$\theta_{\mathrm{max}}$=60 deg. See section 3 for full explanations about
the parameter $\theta_{\mathrm{max}}$. Blue and red crosses show
the error-bars for Auger South and Auger North respectively; the dotted the
expected level for a full uniform sky coverage.}
\label{invmll}
\end{figure}

\subsection{Discussion}

One might ask the question of the pertinence of measuring an angular
power spectrum on a partial region of the celestial sphere and ask
what is the link between such a {\em local} power spectrum and the
{\em global} one. In fact, all of this discussion is linked to spatial stationarity of
the random field. If the field is spatially stationary, then, a partial part of the sky 
if a fair sample and can allow to recover all modes provided of course that the matrix $M_{\ell
\ell^\prime}$ can be inverted, which is the case when the portion of
the sky covered is larger than $\simeq 40\%$ as shown on Fig.~\ref{invmll}. 
If the sky is not spatially stationary (which can be seen by comparing power 
spectra in various partial regions of the sphere), then our procedure cannot be applied and
the recovered power spectrum is of course not valid for the whole sphere. 
In that case anyway, the non stationarity of the sky makes the notion of a single power spectrum on
the sphere totally irrelevant and our procedure allows one to
construct power spectra for different regions provided the fact that
they are spatially stationary enough. We therefore see that in any case,
computing a {\em local} version of the angular power spectrum is
interesting as it provides a {\em global} information when it is
meaningful (spatially stationary sky) or a {\em local} one if relevant.

There are a lot of theoretical motivations to detect large scale patterns
around 1 EeV as well as at higher energies. Magnetic fields or relative motion 
of the observer with respect to cosmic ray rest frame are natural mechanisms
that lead to low-order moments. As these low-order poles characterize
properties over the whole sphere, the condition of spatial stationarity 
should be naturally achieved in those cases, and local power spectra should
show the same patterns whatever the observed portion of the sky. On the other
hand, at higher poles, and if sources are located in large scale structure such as 
the Virgo cluster or the Super Galactic Plane, we expect that power spectra obtained 
from different regions of the sky covering or not those regions will show different 
characteristics. Again, spatial stationarity would be achieved within each region or along 
superstructures and the power spectrum obtained in those cases will give us informations on the sources 
distribution and magnetic fields within those structures. Of course, there is no way to completly 
describe the sky without looking at it. If a spectacular source or set of sources is present somewhere,
the only way to know about it is to achieve (with several observatories) full sky coverage.

The formalism of the angular power estimation is up to now the tool the
most sensitive to search for tiny anisotropies over the sphere, provided
the fact that the detector doesn't smear out the arrival directions more
than the scale $1/\ell$. 
One can imagine that the use of this observable in the next future
will be of great help in order to bring strong constraints on the
UHECRs models of production and propagation.

\section{The Pierre Auger Observatory}

The Pierre Auger Observatory\footnote{Named after the French 
physicist Pierre Auger (1899-1993) who discovered the Extensive Air Showers.} 
is the first of a new generation of
detectors specially dedicated to the highest-energy cosmic rays.
Large area ground based detectors do not observe the incident cosmic 
rays directly but the Extensive Air Showers (EAS), a very large cascade 
of particles, that they generate in the atmosphere. All experiments 
aim to measure, as accurately as possible, the direction of the primary 
cosmic ray, its energy and its nature. There are two major techniques 
used. One is to build a ground array of sensors spread over a large area,
to sample the EAS particle densities on the ground. The other
consists in studying the longitudinal development of the EAS by detecting
the fluorescence light emitted by the nitrogen molecules which are 
excited by the EAS secondaries.

The Auger Observatory~\cite{auger} combines both techniques. The 
detectors are designed to be fully efficient for showers above 10~EeV, 
with a duty-cycle of 100\% for 
the ground array, and 10 to 15\% for the fluorescence telescopes. The 
1600 stations of the ground array are cylindrical \v{C}erenkov tanks 
of 10~m$^2$ surface and 1.2~m height filled with filtered water; they 
are 1.5 km spaced on a triangular grid. 

The Auger observatory covers both hemisphere, one site in the South is presently under
construction in Argentina and will be completed at the end of 
2005. The Northern site construction should start soon after. Once fully 
completed in 2008, the Auger Observatory will be covering 
a surface of 2$\times $3000~km$^2$ and will provide 
unprecedented statistics. With a total aperture of more than 14000~km$^2\cdot$sr,
and with an integral cosmic ray intensity above 10 EeV of approximately 
0.5/(km$^2\cdot$sr$\cdot$yr), 
the Auger Observatory should detect every year of the order of 7000 
events above 10~EeV and 70 above 100~EeV (assuming a $E^{-3}$ dependence
of the spectrum). The angular resolution of the
surface detector alone is believed to be of the order of 1 degree, and
can be improved to less than 1 degree in the case of the so-called
hybrid events, that is, events detected by both the surface detector
and the fluorescence detector~\cite{auger}. 

\begin{figure}[!ht]
\centering\epsfig{file=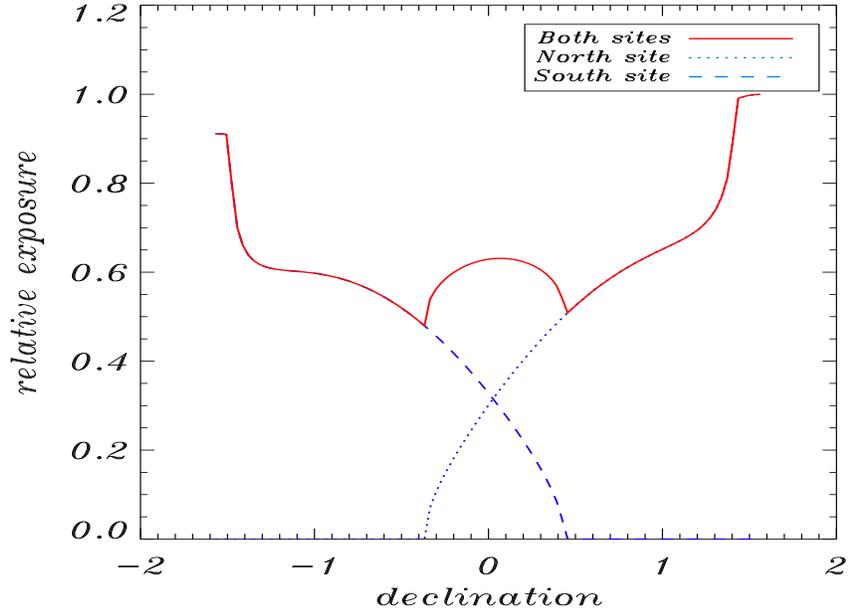,width=13cm,height=9cm}
\caption{The declination dependence of the Auger Observatory relative
exposures. The Southern and Northern sites are indicated separately by
dots, whereas the combined exposure is in solid line.}
\label{auger-expos}
\end{figure}

A full-time operation of the surface detector means that there is
no exposure variation in sideral time and therefore constant exposure
in right ascension. For such a detector located at latitude $a_0$ and
fully efficient for cosmic rays arriving with zenith angles $\theta$
less than some maximal value $\theta_{\mathrm{max}}$, the exposure
is only a function of declination $\delta$~\cite{sommers}~: \[
W(\vec{n}) \propto \cos{(a_0)}\cos{(\delta)}\sin{(\alpha_{\mathrm{m}})}  
+ \alpha_{\mathrm{m}}\sin{(a_0)}\sin{(\delta)}
\]
where $\alpha_{\mathrm{m}}$ is given by~:\[
\alpha_{\mathrm{m}} = \left\{ \begin{array}{cc}
0&\mathrm{if}\,\,\xi>1\\ \pi&\mathrm{if}\,\,\xi<-1\\ \cos^{-1}{(\xi)}&\mathrm{otherwise}\\
\end{array}\right.
\]
and~:\[
\xi=\frac{\cos{(\theta_{\mathrm{max}})}-\sin{(a_0)}\sin{(\delta)}}{\cos{(a_0)}\cos{(\delta)}}
\]
Fig.~\ref{auger-expos} shows the resulting declination dependence
for the two sites of the Auger Observatory located at latitude 
$a_0=+39$~deg. for the Northern one, and $a_0=-35$~deg. for the Southern
one. The cut angle $\theta_{\mathrm{max}}$ is chosen at 60 deg. Also
shown is the combined exposure, which will completely cover the celestial
sphere but will not be uniform.

\begin{figure}[!ht]
\centering\epsfig{file=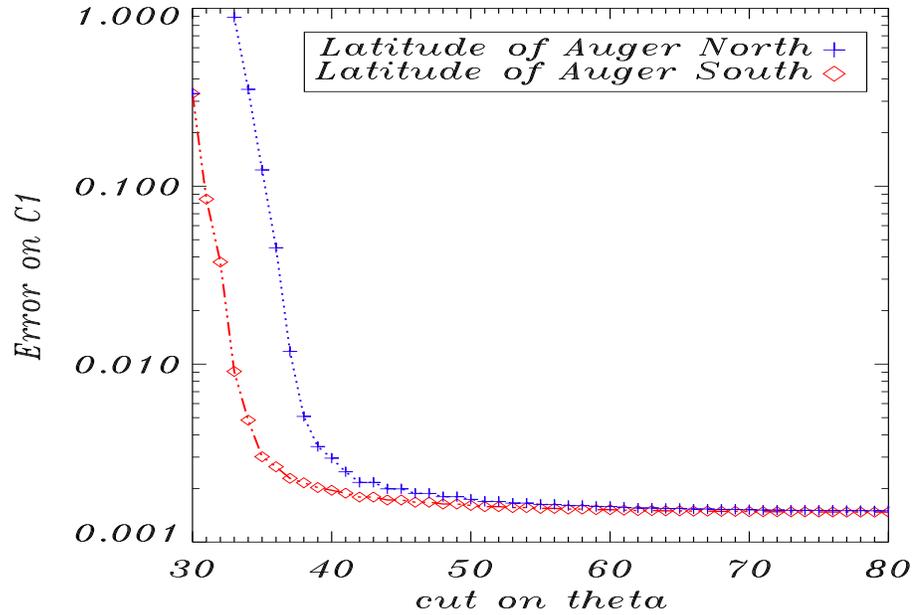,width=13cm,height=9cm}
\caption{Amplitude of the $C_{1}^{\mathrm{exp}}$ error bars as function
of $\theta_{\mathrm{max}}$ for the two Auger sites latitudes, computed
with a set of 7000 events. From $\theta_{\mathrm{max}}=50$ degrees, results 
on the $C_{\ell}^{\mathrm{exp}}$ error bars are not sensitive to the choice 
of $\theta_{\mathrm{max}}$.}
\label{cut-on-theta}
\end{figure}

In order to show that analysis are not
sensitive to the choice of $\theta_{\mathrm{max}}$ for the two
latitudes we are considering, Fig.~\ref{cut-on-theta} plots, for a set
of 7000 events, the variation of the first multi-pole
$C_{1}^{\mathrm{exp}}$ error bars as function of
$\theta_{\mathrm{max}}$.  Clearly, for both sites from 50 deg. and on, the
error bars are stable with respect to the 
$\theta_{\mathrm{max}}$ parameter.

\section{Predicted constraints on isotropy with the Auger detector}

\begin{figure}[!ht]
\centering\epsfig{file=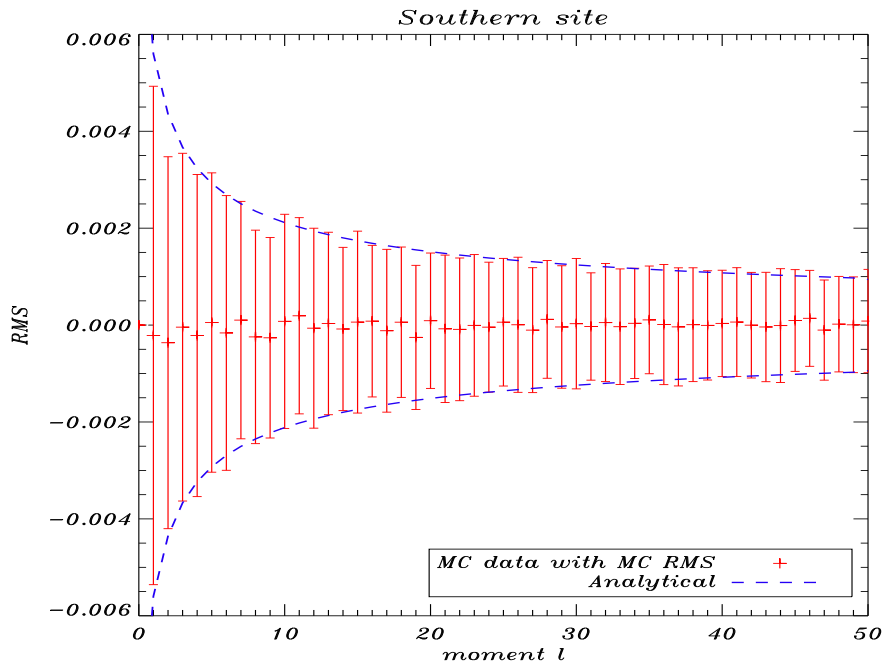,width=12cm,height=9cm}
\caption{Comparison between the analytical estimation of the bias and of
the error-bars and the simulated ones of our angular power spectrum 
estimate. The coverage of the experiment is assumed to be the Auger
Southern one, with a duration of 1 year data taking. We consider only
statistics beyond 10 EeV. Clearly, the analytical computation perfectly
reproduces the properties of the Monte-Carlo simulation.}
\label{augersud}
\end{figure}

In this section, we choose to deal with events with energy beyond 10 EeV, 
where the required fully efficient cosmic rays detection is satisfied
by the Auger Observatory for a large range of $\theta_{\mathrm{max}}$.
The total number of events $N_{\mathrm{tot}}(E)$ being detected with
the Auger arrays on the whole sphere is approximately 
$N_{\mathrm{tot}}(E>10\mathrm{\,EeV}) = 7000$ per year. 
We consider here the number of events $N_{\mathrm{ev}}$ for
an integration time $T$, weighted by the covered fraction of the sky~: \[
N_{\mathrm{ev}} = N_{\mathrm{tot}}(E>10\,\mathrm{EeV})\times T\times f_1
\]
To check our estimate of the bias and of the error bars, we
simulated 100 times $N_{\mathrm{ev}}$ events on a uniform
(zero power spectrum) sky with the Southern Auger site coverage.
We then reconstructed the power spectrum using the relations given 
in the previous sections. Fig.~\ref{augersud} shows the perfect agreement 
between the Monte-Carlo simulations and the analytical predictions for
both the bias (the bias has been subtracted and the resulting values
are indeed centered on zero) and the error bars.
This plot is performed for an integration of 1 year of data taking.

\begin{figure}[!ht]
\centering\epsfig{file=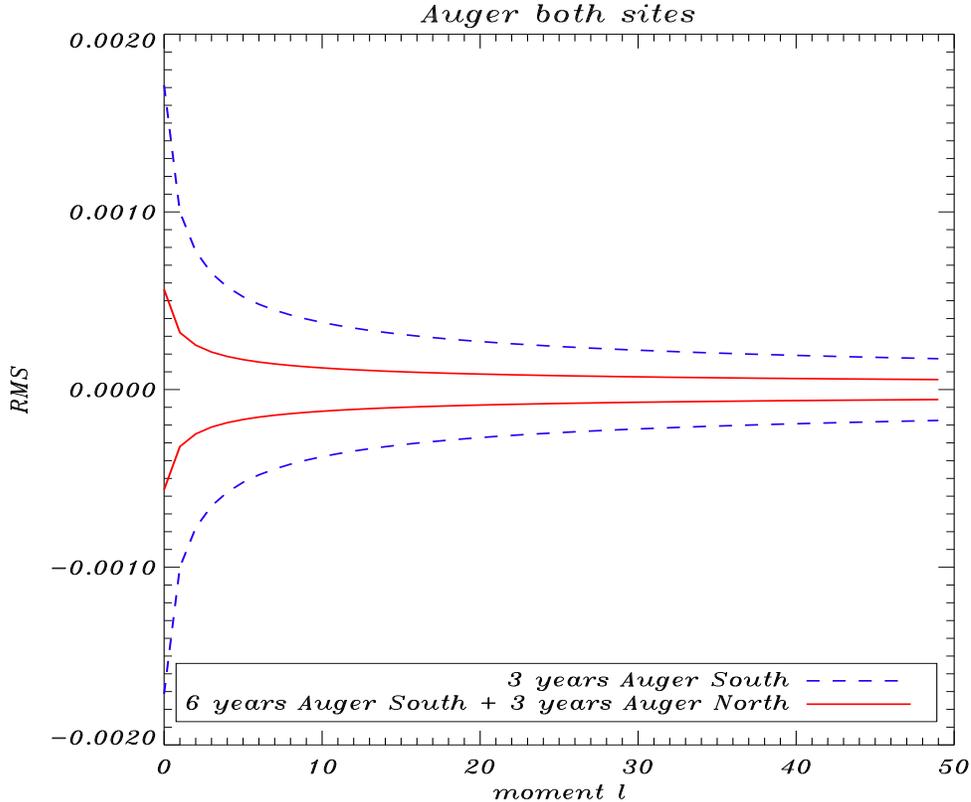,width=13cm,height=11cm}
\caption{Constraints on an isotropic distribution of cosmic rays
beyond 10 EeV for 3 years of data taking of the Southern site only
(dashed line), and for 6 years of data taking of the Southern site
+ 3 years of the Northern one (continuous line). These error-bars
are estimated from the analytical computation.}
\label{augernordsud}
\end{figure}

The enhancement of the number of events with the Northern site
allows for more stringent constraints on an isotropic distribution
as can be seen on Fig.~\ref{augernordsud}. We have imposed 3 years 
of data taking for the Northern site, and 6 years for the Southern 
one (3 years for the only Southern site + 3 years for both the
Southern and Northern sites). With such statistics, it becomes
possible to test the isotropy hypothesis with an accurate precision.

\section{Sensitivity of Auger to a dipole}

In order to check for the efficiency of our mode deconvolution, we
simulated events following a uniform distribution plus a dipole of 10 \% amplitude. 
In practice, from
the $C_{\ell}$ input spectrum, we generated the $a_{\ell m}$ 
coefficients with uniform random phases. We then transformed these 
coefficients into a sky map using an inverse harmonic transform in order 
to have a realization of the $\Delta$ random field. We then multiplied 
this map by the required coverage of the sky, and drew
the number of events falling in each pixel using a Poisson law
with average proportional to the this map. The exact position of 
each event within the pixel is drawn uniformly. All these steps rely
heavily on the software provided along with the Healpix pixellisation
scheme~\cite{healpix}. For each Monte-Carlo realization of a set of events 
for a given input power spectrum, we then extract the pseudo power
spectrum by expanding the map onto the spherical harmonics basis (using the
{\tt anafast} routine) and then apply the deconvolution to reconstruct the 
unbiased and unmixed power spectrum. We did the simulations for the 
Southern site only and for both sites with the same integration of data
taking than in the previous section.

\begin{figure}[!ht]
\centering\epsfig{file=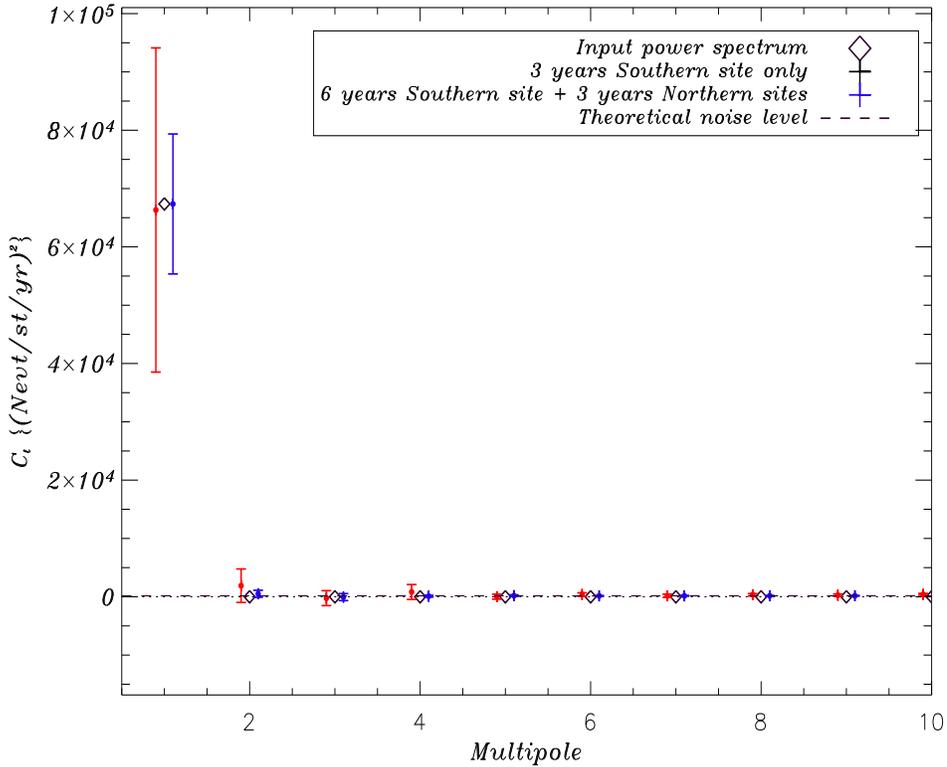,width=13cm,height=11cm}
\caption{Reconstructed power spectrum (expressed here in 
[Number of events/year/str]$^2$ )in case of a pure dipole input
sky. The diamonds show the input power spectrum. 
In red is shown the reconstructed power spectrum with the
only Southern site of the Auger Observatory, for a duration of 3
years of data taking; whereas in blue is shown the reconstructed
power spectrum with both sites, for a duration of 6 years for the 
Southern site and 3 years for the Northern one (as in previous
section).}
\label{dipole}
\end{figure}

The result of the Monte-Carlo are shown on Fig.~\ref{dipole}. The 
reconstructed power spectrum is in agreement with the input one,
and excludes the isotropic distribution at the 2.5 $\sigma$ level
using the Southern site only. Obviously, the improvement
of the reconstructed power spectrum is clear with the 3 years data
taking of the Northern site added to the analysis. As a
consequence, the error-bars of the quadrupole are strongly reduced.
The same procedure can be applied to any input power spectrum.

\section{Conclusions}
We showed that in the general case of a varying and incomplete exposure on the sky, the true power
spectrum of the cosmic ray sources distribution can be recovered. This
result is not new in itself as it was introduced a few years ago in
the framework of CMB data analysis~\cite{hivon}. Its application to
cosmic ray data is however new and might open new possibilities as the
general feeling up to now was that no $C_\ell$ power spectrum can be
reconstructed without a complete sky coverage. The power spectrum that
our procedure allows to recover is equivalent to the full sky one if
the anisotropies in the arrival directions of the cosmic rays are well
modeled by a spatially stationary random field on the sphere. If this is not
the case, the recovered power spectrum is still valid, but only for
the region that was used to determine it. Anyway in the non
spatially stationary case, different power spectra are required in different
regions of the sky and our approach is still relevant.  Additionally,
we have analytically solved the calculation of the bias and of the variance
introduced by the finite sampling of the sky in the general case of a
varying and eventually incomplete exposure. Using the deconvolution
proposed here, any cosmic ray dataset will be usable for anisotropy
determination purpose, provided the fact that the arrival directions
and coverage map are known within reasonable precision.

\section*{Acknowledgments}
The authors wish to thank amically Etienne Parizot for profound
discussions and Paul Sommers for fruitful objections and discussions.

\end{document}